\documentclass[aps,prl,reprint,groupedaddress]{revtex4-2}

\usepackage{graphicx}
\usepackage{dcolumn}
\usepackage{bm}
\usepackage{xcolor}

\begin{document}


\title{Identifying and anticipating the threshold bifurcation of a complex laser with permutation entropy}


\author{Juan Gancio}
 \email{Contact author: juan.gancio@upc.edu}
\author{Cristina Masoller}%

\affiliation{Universitat Politècnica de Catalunya, Departament de Fisica, Rambla Sant Nebridi 22, Terrassa 08222, Barcelona, Spain.
}%

\author{Mathias Marconi}

\affiliation{Université Côte d’Azur, CNRS, Institut de Physique de Nice, 06200, Nice, France}%


\date{\today}

\begin{abstract}
The permutation entropy (PE) is a statistical indicator that allows to quantify the complexity of a signal.
Here we show that it is able to {identify and anticipate} the threshold bifurcation of a 
complex laser, where 
thousands of modes compete for gain at the onset of lasing.
In our experimental setup, the cavity roundtrip time is several orders of magnitude longer than the temporal resolution of the detection system, which enables a high statistical sampling of the intensity dynamics per roundtrip. We show that the permutation entropy experiences a clear decrease far below the threshold and reaches a sharp minimum 
at the threshold bifurcation point, which reveals an abrupt increase of the temporal correlations. The evolution of the entropy is compared with standard quantifiers of approaching bifurcations. While lag-1 autocorrelation gradually grows as the threshold is approached, PE shows a steep decrease that captures the emergence of nonlinear correlations and thus, it allows a clearer identification of the threshold.


\end{abstract}

\maketitle


The laser threshold marks the transition from spontaneous to stimulated emission, and is a fascinating phenomenon that has intrigued and attracted a lot of attention since pioneering works in the late 60s and early 70s \cite{Risken65,Risken67,Arecchi71,Arecchi73}.  
The threshold transition simultaneously entails a strong increase of the output power, a strong increase of the temporal and spatial coherence of the emitted light and a change in the statistics of the photon number fluctuations from thermal to poissonian distributions \cite{Loudon}. 

Motivated by the development of integrated laser technologies, the analysis of photon statistics and threshold characterization in small laser systems, where only one or a few modes compete for gain at the onset of lasing, has been an intense topic of interest \cite{Ulrich07,Wiersig09,Marconi18,Wang20}.  However, little is known about laser threshold characterization in highly multimode lasers. 
Multimode laser systems are today receiving considerable attention for their potential in light structuration and manipulation for a wide variety of applications \cite{Cristiani22}, and for the analysis of novel spatio-temporal phenomena \cite{Shen23,Bittner25}. The complexity which arises due to
nonlinear interactions of a large number of modes is still
far from being understood \cite{giacomelli_prl_2018,cao_nrp_2019,cao_prl_2022,Cristiani22,Shen23,Carroll23,Bittner25}.

From the point of view of nonlinear dynamics, at threshold, the laser intensity undergoes a transcritical bifurcation from a off-solution (below threshold) to a non-zero intensity solution (above threshold) \cite{Erneux_Glorieux_2010}.
At the onset of a bifurcation, a dynamical system generically  experiences an asymptotic growth of its relaxation time, a phenomenon known as the “Critical Slowing Down” (CSD) \cite{Kramer1985,Scheffer09,Scheffer12}.
It was shown that CSD can cause a delay in the laser turn-on when there is a rapid sweep of the pump parameter \cite{Tredicce04}, and it has been measured
only recently in a laser with modulated losses \cite{Marconi24}.
As CSD indicates a bifurcation and, more generally, a ``tipping point'' in a dynamical system \cite{sebastian}, its anticipation through the analysis of statistical indicators such as increase of variance and lag-1 autocorrelation is receiving considerable attention \cite{dakos_2012,veraart_nature_2012, Klaus_2024, Zhang_prl_2024}.

However, these indications may fail to detect bifurcations or critical transitions before they occur \cite{Klaus_2019,Marconi20,giulio}, and it is therefore essential to develop advanced tools able to detect weak signatures of approaching bifurcations. 

Data recorded during turn-on transition to lasing offers an opportunity to test novel tipping indicators, in very noisy data. However, the main challenge is provided by the time scales of the laser, because the build-up is very fast and even with an extremely fast detection system, the short duration of the build-up prevents the early detection of CSD. To overcome this limitation, here we use a long cavity that enables high statistical sampling of the intensity dynamics per roundtrip. During the transition to lasing, the intensity of such a highly multimode laser displays complex structures that form within a cavity roundtrip due to the fast competition between many longitudinal modes \cite{Roche23}. Therefore, appropriate analysis tools are needed to fully capture the complexity of the intensity fluctuations at threshold.

Ordinal analysis and permutation entropy \cite{bandt2002permutation,bandt2005ordinal} are popular nonlinear analysis tools used to identify bifurcations \cite{cao_2004,masoller_2015} and have been used to characterize complex signals in different systems \cite{rosso_prl,
masoller_epl_2022}, including chaotic external-cavity lasers \cite{zunino2011characterizing,soriano2011time,aragoneses2014unveiling} and long-cavity lasers \cite{aragoneses2016unveiling,min_oe_2021}. 
Ordinal analysis defines symbols --known as ordinal patterns-- according to the relative values of $L$ datapoints in the time series (the method is explained in detail in Sec. I of the {\it Supplementary Information}\cite{sup_mat}). The method is computationally efficient and robust to noise \citep{bandt2002permutation}. 
The use of non-consecutive data points to define the symbols 
adds versatility to the method, since it allows to select the temporal scale of the analysis \citep{bandt2005ordinal,soriano2011time,aragoneses2016unveiling}. An extension of this method to two-dimensional data \cite{haroldo} was recently used to analyze speckle patterns, whose contrast quantifies the degree of spatial coherence of the laser beam \cite{tirabassi2023permutation}; however, this method was not able to provide a clear identification of the threshold. Here we demonstrate that the permutation entropy is able to detect the approaching turn-on, and accurately identifies the threshold. 

\begin{figure}[tb] 
    \centering
    \includegraphics[width=0.99\linewidth]{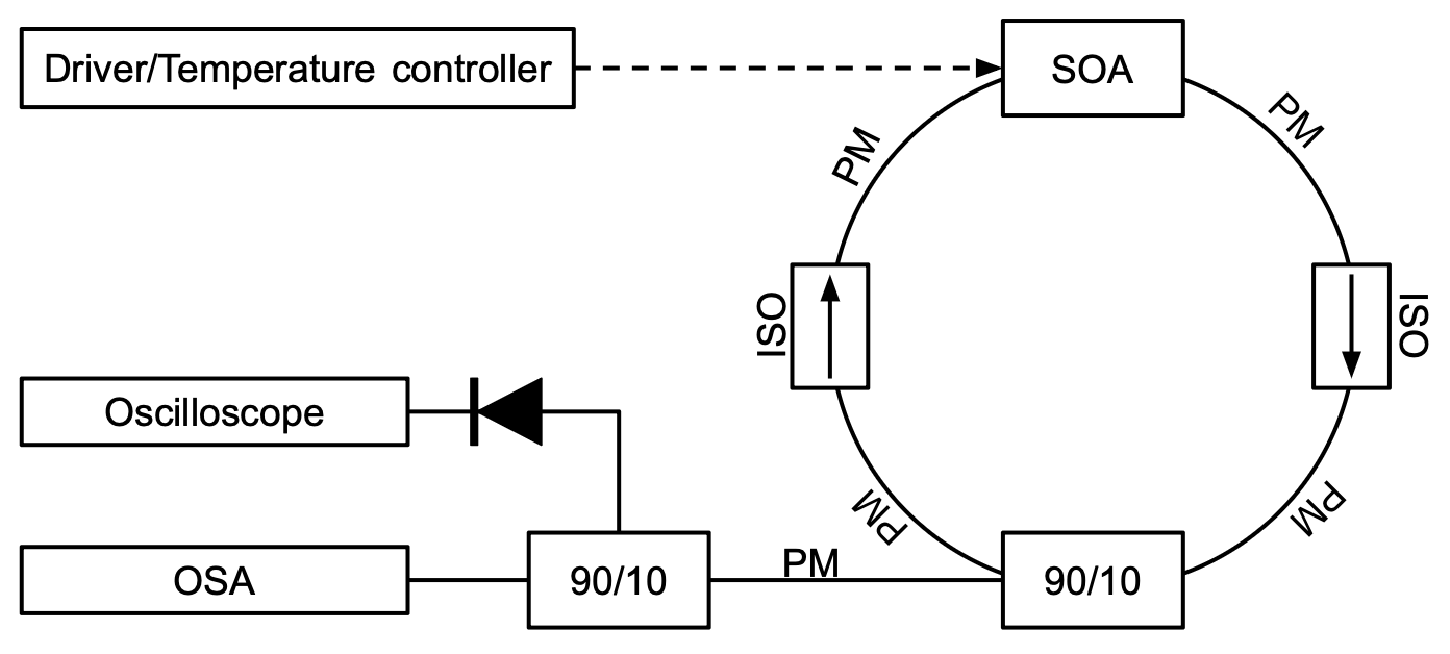}
    \caption{Experimental setup: a
semiconductor ring laser consisting of a semiconductor optical amplifier (SOA) that is contained in a loop of polarization-maintaining (PM) fibers. Optical isolators (ISO) are used to prevent spurious reflections inside the cavity. 90 $\%$ of the light is extracted from the cavity using a 90/10 fiber splitter and coupled into our detection system consisting of a photodetector, an oscilloscope, and an optical spectrum analyzer (OSA). }
    \label{fig:1}
\end{figure}

\begin{figure*}[tb] 
    \centering
    \includegraphics[width=.9\linewidth]{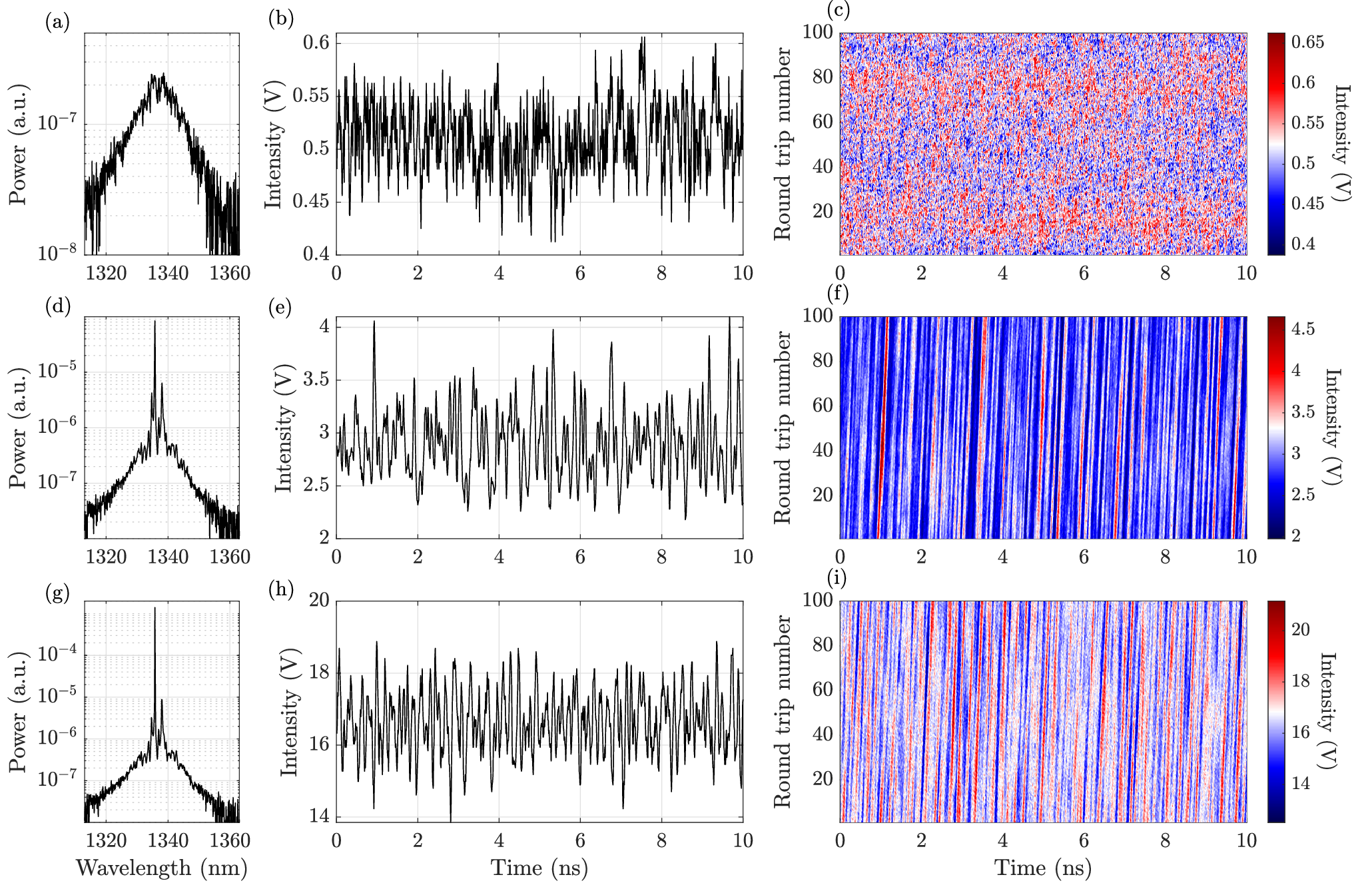}
    \caption{Optical spectra (left column), temporal evolution of the laser intensity during 10~ns (central column), and 2D maps of the laser intensity (in color code) zoomed in 10~ns segments over $100$ roundtrips. The color code is normalized such that in each panel, red (blue) represents large (low) intensity values. In panels (a-c) the pump current is below threshold ($48$~mA), in (d-f), is at threshold ($51$~mA), and in (g-i), is above threshold ($55$~mA).}
    \label{fig:2}
\end{figure*}

The experimental setup, shown in Fig.~\ref{fig:1}, consists of 
a semiconductor ring laser in which a semiconductor optical amplifier (SOA) emitting at 1340 nm is contained in a fiber loop. In order to force the laser to operate unidirectionally and to avoid spurious reflections inside the resonator, two optical isolators are inserted. Polarization-maintaining fibers are used to suppress polarization dynamics inside the resonator. The laser cavity length is $6.76$~m, corresponding to a $33.8$~ns roundtrip time.



A fiber splitter couples 90 \% of the light into the detection setup that simultaneously monitors the average optical spectrum and the real-time intensity dynamics with a 10 GHz photodetector connected to a 33 GHz, 100 GSamples/s digital oscilloscope.

The main advantages of our setup for studying the intensity dynamics at threshold are: i) the spontaneous emission 
of the semiconductor gain medium below threshold is guided into the fiber, which therefore optimizes the number of photons 
collected for detection, ii) the SOA's gain is very high (30 dB) which allows to obtain lasing even under 
more than 90 \% losses in the cavity. As a consequence, a large portion of the light can be sent to the detection which allows having enough signal to observe fast temporal fluctuations even below the threshold. iii) The fiber-loop cavity is long enough, 
such that the roundtrip time is much longer than the temporal resolution of the detection system (about 50 ns vs. 10 ps respectively). 

\begin{figure}[tb] 
    \centering
    \includegraphics[width=0.9\linewidth]{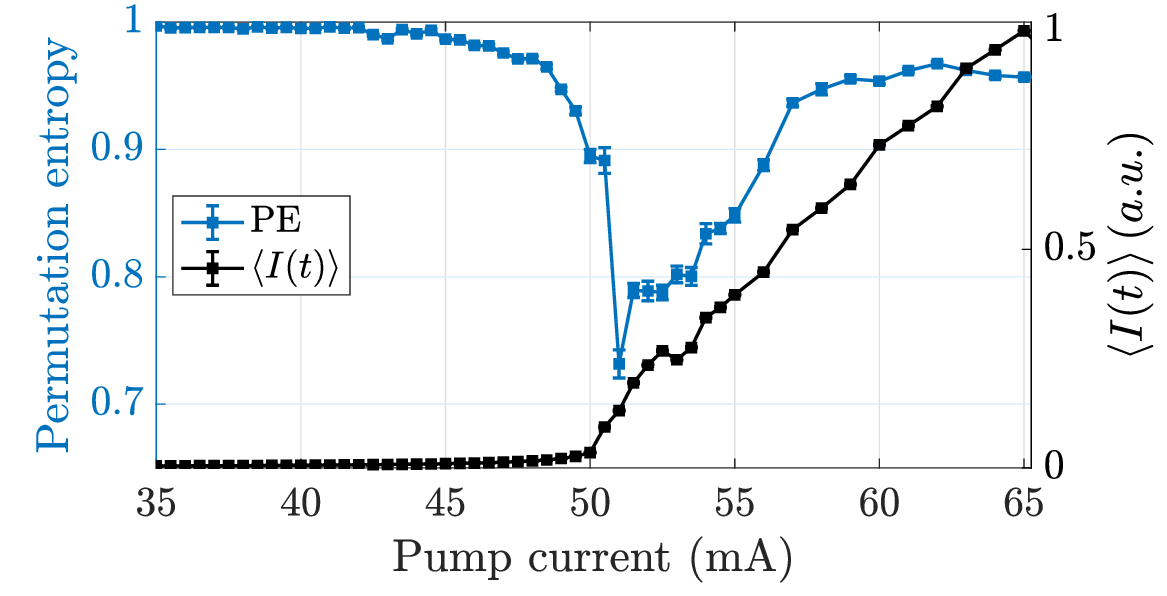}   
    \caption{Permutation entropy (blue) calculated from the probabilities of ordinal patterns defined by four consecutive data points (i.e., $L =4$ and $\tau=1$) and mean value of the output intensity, $\left<I(t)\right>$ (black), vs. the pump current. Error bars represent $1$ standard deviation around the mean obtained from the analysis of $100$ non-overlapping windows, although they are not visible (except at threshold) because their size does not exceed the size of the symbol that shows the mean value.}
    \label{fig:3}
\end{figure}


Thus, our experimental setup allows us to measure the intensity fluctuations with a sampling of about 1000 points within a roundtrip. 
We recorded intensity time traces with 
$4\times10^7$ data points sampled at $100$~GS/s. The time traces were recorded for $51$ values of the pump current, from $35$~mA to $65$~mA,  
which cover the range from below to above the threshold. The data collected is publicly available in \cite{data}.

Figure~\ref{fig:2} displays typical optical spectra and intensity time traces recorded below, at and above threshold.
Below threshold (Fig.~\ref{fig:2}a) the spectrum displays a 
Gaussian shape and a broad linewidth of about 20 nm. The ratio between the gain linewidth and the cavity free spectral range is of the order of $10^5$, hence a huge quantity of modes compete for the gain. At threshold, a sharp peak emerges 
(Fig.~\ref{fig:2}d), which increases at higher pump current (Fig.~\ref{fig:2}g).

The evolution of the laser intensity during 10 ns is shown 
in the central column (Figs. \ref{fig:2}(b), \ref{fig:2}(e) and \ref{fig:2}(h)). In such short time interval, the intensity fluctuations appear fully irregular and the only difference in the three panels is in the vertical scale.

However, using the value of the roundtrip time, we can represent the evolution of the intensity dynamics over many round-trips in the form of a 2D map \cite{Giacomelli96,masoller97,aragoneses2016unveiling,Roche23,Bartolo23}. We consider $N$ segments of the intensity time series,  each of $T$ data points that correspond to the roundtrip time $T dt=33.8ns$ (here $dt=10ps$ is the sampling time), such that $NT\le M$, where $M$ is the total number of data points in the time series. Then, we plot the intensity value in color code on a 2D map, where the vertical axis is the roundtrip number ($i=1\dots N$, temporal dimension) and the horizontal axis is the time ($t=1\dots Tdt$). Therefore, the 2D map is a stroboscopic representation of the temporal evolution of the intensity, where the stroboscopic period is the cavity roundtrip time.

In the 2D maps shown in Fig.~\ref{fig:2} we represent segments of 10 ns over 100 roundtrips. We observe that below threshold (Fig.~\ref{fig:2}c) the fluctuations appear uncorrelated.  At threshold (Fig.~\ref{fig:2}f), we note the formation of persistent and highly contrasted structures. 
Surprisingly, the lifetime of some structures can be extremely long, such as the one appearing at 1 ns, which is of the order of 100 roundtrips (a zoomed trace of that structure in presented in Fig. 1 of the Supplementary Information \cite{sup_mat}, SI). We assume that these structures are caused by the complex beatings of laser cavity modes. At higher pump current (Fig.~\ref{fig:2}f), more structures appear but their contrast as well as their lifetime visibly decrease. 

\begin{figure}[tb] 
    \centering
    \includegraphics[width=0.9\linewidth]{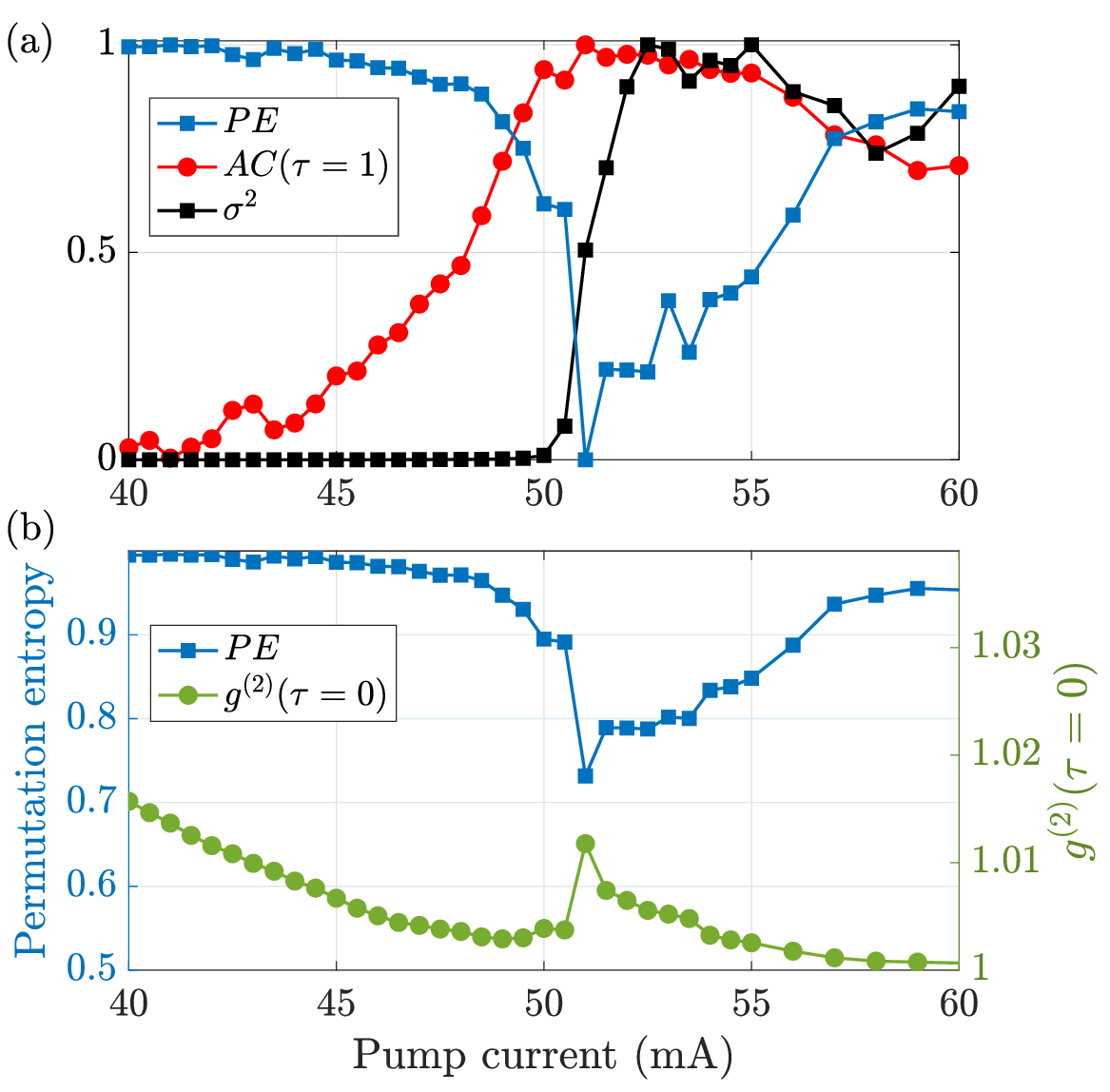}
    \caption{(a) Permutation entropy (blue) calculated from the probabilities of ordinal patterns defined by four consecutive data points (i.e., $L =4$ and $\tau=1$), lag-1 autocorrelation (red), and  variance, $\sigma^2$ (black), of the laser intensity. The values of $PE$, $AC$ and $\sigma^2$ are normalized between $0$ and $1$. (b) Permutation entropy (blue) and $g^{(2)}(\tau=0)$ (green).}
    \label{fig:4}
\end{figure}

To characterize the complexity of the intensity dynamics in different time-scales, we apply the 
ordinal method, which is based on defining symbols, known as ordinal patterns (OPs), which encode information about the temporal order of the relative values of $L$ data points in a time series that are separated by a lag $\tau$ (see section I of the SI). Then, the symbols' probabilities, known as ordinal probabilities, are estimated from the frequencies of occurrence of the different symbols in the time series. Shannon's entropy is computed from these probabilities, 
\begin{equation}
H=-\frac{1}{\log\left(L!\right)}\sum_{j=1}^{L!}p_j\log\left(p_j\right),\end{equation} 
which is known as permutation entropy (PE) and is normalized such that, when all the symbols' probabilities are equal, $H=1$, while when only one symbol is present in the time series, $H=0$.  

Figure~\ref{fig:3} displays the variation of the permutation entropy with the pump current, together with the Light output--Input (LI) curve, calculated for $L =4$ and $\tau=1$ (i.e., four consecutive data points),  
which represents the average emitted intensity, $\left<I(t)\right>$, vs. the pump current. 
In the LI curve we observe that the lasing transition occurs at a pump current of 50 mA. 
\begin{figure}[tb] 
    \centering
    \includegraphics[width=0.9\linewidth]{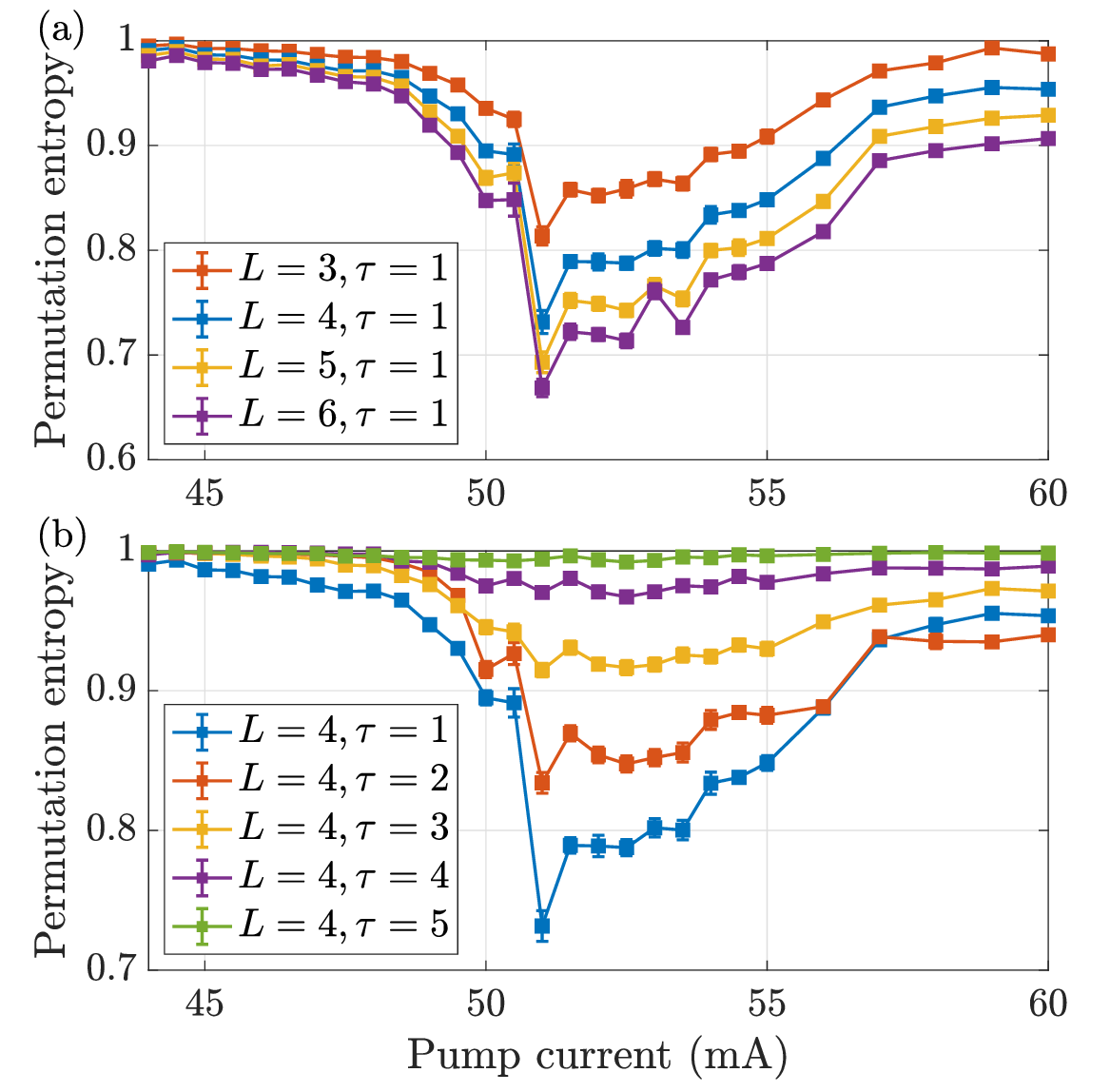}
    \caption{Permutation entropy for different word lengths ($L$), and delays ($\tau$). In panel (a), the time traces' lengths used to obtain the PE are scaled by a factor of $L!$ in order to estimate properly the probabilities of the increasing number of symbols. Error bars represent one standard deviation around the mean obtained from the analysis of $100$ non-overlapping segments of the intensity time-series.}
    \label{fig:5}
\end{figure}
In the PE curve, we see that well below the threshold PE is very close to 1, which indicates that the signal in the photodector is (as expected for spontaneous emission) 
uncorrelated noise. From about 42 mA, 
we observe that the PE value starts decreasing, which reveals that the temporal signal is getting more and more ordered. 
The minimum PE value, which indicates maximum strength of temporal correlations, occurs for a pump current of 51 mA. 
We interpret this point as the threshold bifurcation point, where the signal has reached a maximum order. 
Above this point, PE increases and saturates to values close but lower than one, which reveals that correlations persist in the intensity fluctuations well above the threshold (as can be expected due to the residual dynamics caused by mode competition in this complex laser, see section V of the SI). Comparing the evolution of the entropy with the LI curve, we note that PE reaches a minimum value just after the power starts to grow; however, PE in fact anticipates the laser threshold because it begins to decrease much earlier.

To test the robustness of these observations, we performed a second set of experiments using a longer cavity and found qualitatively and quantitatively the same results (see Sec.~III of the SI). 
In the two experiments, the minimum PE value occurs just after the power starts growing, but PE starts decreasing much earlier.


We next compare the variation of the  PE indicator, with the variation of two conventional indicators that have been used as ``early indicators'' of regime transitions in a variety of systems \cite{Scheffer12}: i) the lag-1 autocorrelation $AC=\langle (I(t)-\mu)((I(t+1)-\mu)\rangle/\sigma^2$, where $\mu=\left<I(t)\right>$, and ii) the variance $\sigma^2$, of the intensity fluctuations. 

In Fig.~\ref{fig:4}(a) we observe that AC captures a gradual increase of temporal correlations in the intensity below threshold, and reaches a maximum after the threshold. At difference with PE, 
the threshold is not indicated by a sharp variation, but rather by a smooth maximum around the threshold region. Therefore, both PE and AC capture the increase of correlations below threshold; however, while AC evolution is gradual, PE evolution is more pronounced due to its ability to capture nonlinear correlations. 
Bifurcations in complex stochastic systems are also often marked by an increase of the variance \cite{Scheffer12}, which is a consequence of the CSD phenomenon.
We notice that the threshold transition is marked by a sharp growth of the variance when the pump current is above 50 mA; this behavior occurs in the range of pump power for which correlations are maximum. Therefore, the threshold is neatly identified by the variance. However, while correlations decrease after threshold, the variance does not follow a clear evolution. Therefore, PE provides additional information about the state of the system compared to the analysis of the variance of the fluctuations.


Lag-1 AC and variance are two generic indicators of transitions occurring in complex dynamical systems. In the more specific framework of laser photon statistics, the threshold is often characterized using the evolution of the second-order coherence at zero delay, which, for a classical field is $g^{(2)}(\tau=0) = \left<I^2(t)\right>/\left<I(t)\right>^2 = 1 + \sigma^2/\left<I(t)\right>^2$ \cite{Loudon}.
The value of $g^{(2)}(0)$  allows a distinction to be made between superpoissonian fluctuations below threshold ($g^{(2)}(0) > 1$) and poissonian fluctuations above threshold ($g^{(2)}(0)$ tending to 1). 
In Fig. \ref{fig:4}(b) we observe that below threshold, as expected,
$g^{(2)}(0)$ has a value larger than one.
At threshold, the variance of the intensity abruptly increases (see Fig. 4a), which causes $g^{(2)}(0)$ to display a small kink. Then, $g^{(2)}(0)$ continues decreasing and reaches a value very close to 1 well above the threshold, as expected for the poissonian fluctuations of the laser. Therefore, the strong fluctuations at the onset of the threshold are indicated by the small kink in the $g^{(2)}(0)$. Hence, while variance and $g^{(2)}(0)$ identify the threshold via the increase of the fluctuations, PE provides additional information by quantifying the increase of the correlations in the fluctuations.



Finally, we analyze the robustness of the PE indicator with respect to the parameters $L$ and $\tau$ used to define the ordinal patterns. In Fig.~\ref{fig:5}(a) we see that increasing the length of the pattern, $L$, enhances the effect captured by PE, as PE experiences larger variations that mark the threshold. 
This indicates that, at threshold, complex temporal structures appear in the intensity fluctuations, which longer symbols can properly uncover. 
On the other hand, as shown in Fig.~\ref{fig:5}(b), increasing the lag, $\tau$, between the data points used to define the symbols has the opposite effect, it decreases the PE variation at the threshold crossing. The analysis of the optimal set of parameters, $L$ and $\tau$, that maximize the variation of the PE at threshold is presented in the Sec.~IV of the SI.



To conclude, we have provided a complete analysis of the intensity dynamics at the threshold of a complex laser system. We have shown that PE is able to quantify nonlinear temporal correlations in the intensity fluctuations and anticipate the approaching threshold bifurcation. The indirect indication of CSD provided by PE 
complements that provided by the traditional indicators 
because, in contrast with traditional indicators, PE shows a steep decrease that allows a clear identification of the threshold.

We have shown that its performance is robust to changes in the parameters used to define the ordinal patterns. Moreover, a second set of experiments performed with a longer cavity (presented in the SI) shows an excellent qualitative and quantitative agreement. 
Therefore, our study shows that PE can be used to anticipate and identify the threshold of complex laser systems, where mode competition during the build-up of the laser beam makes difficult to precisely identify the onset of lasing. We propose that our work could be extended to the analysis of many-mode laser systems with high spatio-temporal complexity \cite{cao_prl_2022,Bittner25} and other systems, such as Bose-Einstein condensates of polaritons or photons \cite{Schmitt14, Ozturk23}, to identify approaching bifurcations and transitions.


{\it Acknowledgment} J. G. and C. M. acknowledge partial financial support  of Agència de Gestió d’Ajuts Universitaris i de Recerca (FI AGAUR 2023 and 2021 SGR 00606) and the Institució Catalana de Recerca i Estudis Avançats (ICREA Academia).

\bibliography{apssamp.bib}

\end{document}